\documentclass{SciPost}

\hypersetup{
    colorlinks,
    linkcolor={red!50!black},
    citecolor={blue!50!black},
    urlcolor={blue!80!black}
}

\usepackage[bitstream-charter]{mathdesign}
\usepackage{aas_macros}
\DeclareSymbolFont{usualmathcal}{OMS}{cmsy}{m}{n}
\DeclareSymbolFontAlphabet{\mathcal}{usualmathcal}

\fancypagestyle{SPstyle}{
\fancyhf{}
\lhead{\colorbox{scipostdeepblue}{\bf \color{white} ~SciPost Physics Proceedings }}
\rhead{{\bf \color{scipostdeepblue} ~Submission }}

\fancyfoot[C]{\textbf{\thepage}}
}

\begin{document}

\pagestyle{SPstyle}

\begin{center}{\Large \textbf{\color{scipostdeepblue}{
Towards a foundation model for astrophysical source detection: An End-to-End Gamma-Ray Data Analysis Pipeline Using Deep Learning\\
}}}\end{center}

\begin{center}\textbf{
Judit Pérez-Romero\textsuperscript{1$\star$},
 Saptashwa Bhattacharyya\textsuperscript{1}, 
 Sascha Caron\textsuperscript{2, 3},
 Dmitry Malyshev\textsuperscript{4},
 Rodney Nicolas\textsuperscript{2},
 Giacomo Principe\textsuperscript{5},
 Zoja Rokavec\textsuperscript{6},
 Roberto Ruiz de Austri\textsuperscript{7},
 Danijel Skočaj\textsuperscript{8},
 Fiorenzo Stoppa\textsuperscript{9},
 Domen Tabernik\textsuperscript{8}, and
 Gabrijela Zaharijas\textsuperscript{1}
}\end{center}

\begin{center}
{\bf 1} Center for Astrophysics and Cosmology, University of Nova Gorica, Vipavska 11c, 5270 Ajdovščina, Slovenia
\\
{\bf 2} Nikhef, Dutch National Institute for Subatomic Physics, Science Park 105, 1098 XG Amsterdam, The Netherlands
\\
{\bf 3} High Energy Physics, Radboud University Nijmegen, Heyendaalseweg 135, 6525 AJ Nijmegen, The Netherlands
\\
{\bf 4} Erlangen Centre for Astroparticle Physics, Nikolaus-Fiebiger-Str. 2, Erlangen 91058, Germany
\\
{\bf 5} Dipartimento di Fisica, Università di Trieste, I-34127 Trieste, Italy\\
{\bf 6} Department of Physics, Faculty of Mathematics and Physics, University of Ljubljana, Jadranska ulica 19, 1000 Ljubljana, Slovenia
\\
{\bf 7} Instituto de Física Corpuscular, IFIC-UV/CSIC, Catedrático José Beltrán 2, E-46980 Paterna, Spain
\\
{\bf 8} Faculty of Computer and Information Science, University of Ljubljana, Večna pot 113, 1000 Ljubljana, Slovenia
\\
{\bf 9} Astrophysics sub-Department, Department of Physics, University of Oxford, Denys Wilkinson Building, Keble Road, Oxford, OX1 3RH, UK
\\[\baselineskip]
$\star$ \href{mailto:email1}{\small judit.perez@ung.si}\
\end{center}

\definecolor{palegray}{gray}{0.95}
\begin{center}
\colorbox{palegray}{
  \begin{tabular}{rr}
  \begin{minipage}{0.37\textwidth}
    \includegraphics[width=60mm]{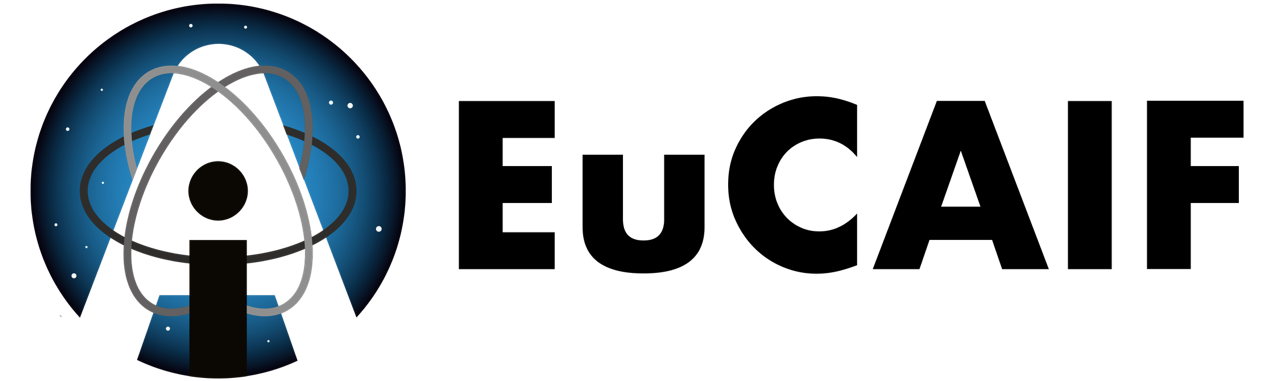}
  \end{minipage}
  &
  \begin{minipage}{0.5\textwidth}
    \vspace{5pt}
    \vspace{0.5\baselineskip} 
    \begin{center} \hspace{5pt}
    {\it The 2nd European AI for Fundamental \\Physics Conference (EuCAIFCon2025)} \\
    {\it Cagliari, Sardinia, 16-20 June 2025
    }
    \vspace{0.5\baselineskip} 
    \vspace{5pt}
    \end{center}
    
  \end{minipage}
\end{tabular}
}
\end{center}

\section*{\color{scipostdeepblue}{Abstract}}
\textbf{\boldmath{%
The increasing volume of gamma-ray data demands new analysis approaches that can handle large-scale datasets while providing robustness for source detection. We present a Deep Learning (DL) based pipeline for detection, localization, and characterization of gamma-ray sources. We extend our AutoSourceID (ASID) method, initially tested with \textit{Fermi}-LAT simulated data and optical data (MeerLICHT), to Cherenkov Telescope Array Observatory (CTAO) simulated data. This end-to-end pipeline demonstrates a versatile framework for future application to other surveys and potentially serves as a building block for a foundational model for astrophysical source detection.
}}

\vspace{\baselineskip}

\noindent\textcolor{white!90!black}{%
\fbox{\parbox{0.975\linewidth}{%
\textcolor{white!40!black}{\begin{tabular}{lr}%
  \begin{minipage}{0.6\textwidth}%
    {\small Copyright attribution to authors. \newline
    This work is a submission to SciPost Phys. Proc. \newline
    License information to appear upon publication. \newline
    Publication information to appear upon publication.}
  \end{minipage} & \begin{minipage}{0.4\textwidth}
    {\small Received Date \newline Accepted Date \newline Published Date}%
  \end{minipage}
\end{tabular}}
}}
}



\section{Introduction}
\label{sec:intro}
Gamma-ray astronomy offers a unique opportunity to study the origin of cosmic rays (CRs) and probe new physics. However, it suffers from major challenges due to the combination of data from different instruments, the diversity of gamma-ray sources, and limitations of current analysis techniques. With nearly one-third of sources in gamma-ray catalogs still unidentified, addressing these shortcomings is essential.

Deep Learning (DL) has shown impressive performance in handling large datasets, with Convolutional Neural Networks (CNNs) being especially suited for vision-related tasks. In astrophysics, such methods have been successfully applied to gamma-ray analyses, including the study of Galactic Center emission \cite{Caron:2017udl} and AutoSourceID \cite{Panes:2021zig} (ASID), the first DL-based tool for automated source detection and classification on gamma-rays.

The case of gamma-ray astronomy exemplifies the broader importance of multi-wavelength approaches for source detection and classification. The integration of multi-wavelength datasets is especially relevant in catalog construction, where cross-matching is performed independently and without a unified framework. Yet, the lack of standardization and the instrument-specific characteristics pose substantial challenges to integrating heterogeneous datasets.

In the works presented in these proceedings, we present our preliminary results of the updated ASID framework on 10 years of \textit{Fermi} Large Area Telescope (LAT) data and on Cherenkov Telescope Array Observatory (CTAO) toy simulations of the Galactic Plane Survey (GP). Finally, we also provide initial indications of the adaptability of the method to be extended to other wavelengths, describing its potential as a first step towards a foundational model for source detection and classification.

\section{\textit{Fermi}-LAT: towards the first DL gamma-ray catalog}
\label{sec:fermi}

Over the last 17 years, \textit{Fermi}-LAT \cite{2009ApJ...697.1071A} has provided an unparalleled contribution to our understanding of the MeV – GeV gamma-ray sky through the production of comprehensive gamma-ray catalogs (see e.g. \cite{Ballet:2020hze}). A central challenge in producing these catalogs is the Galactic diffuse emission (IEM), which dominates the low-latitude gamma-ray sky beyond 50 MeV. Although the LAT collaboration has developed IEM models \cite{Fermi-LAT:2016zaq} to construct these catalogs, the significant model-related uncertainties render them strongly model-dependent and limit the detectability of faint, low-latitude sources, which are crucial for the discovery of new physics.

\begin{figure}[h!]
\centering
\includegraphics[width=\linewidth]{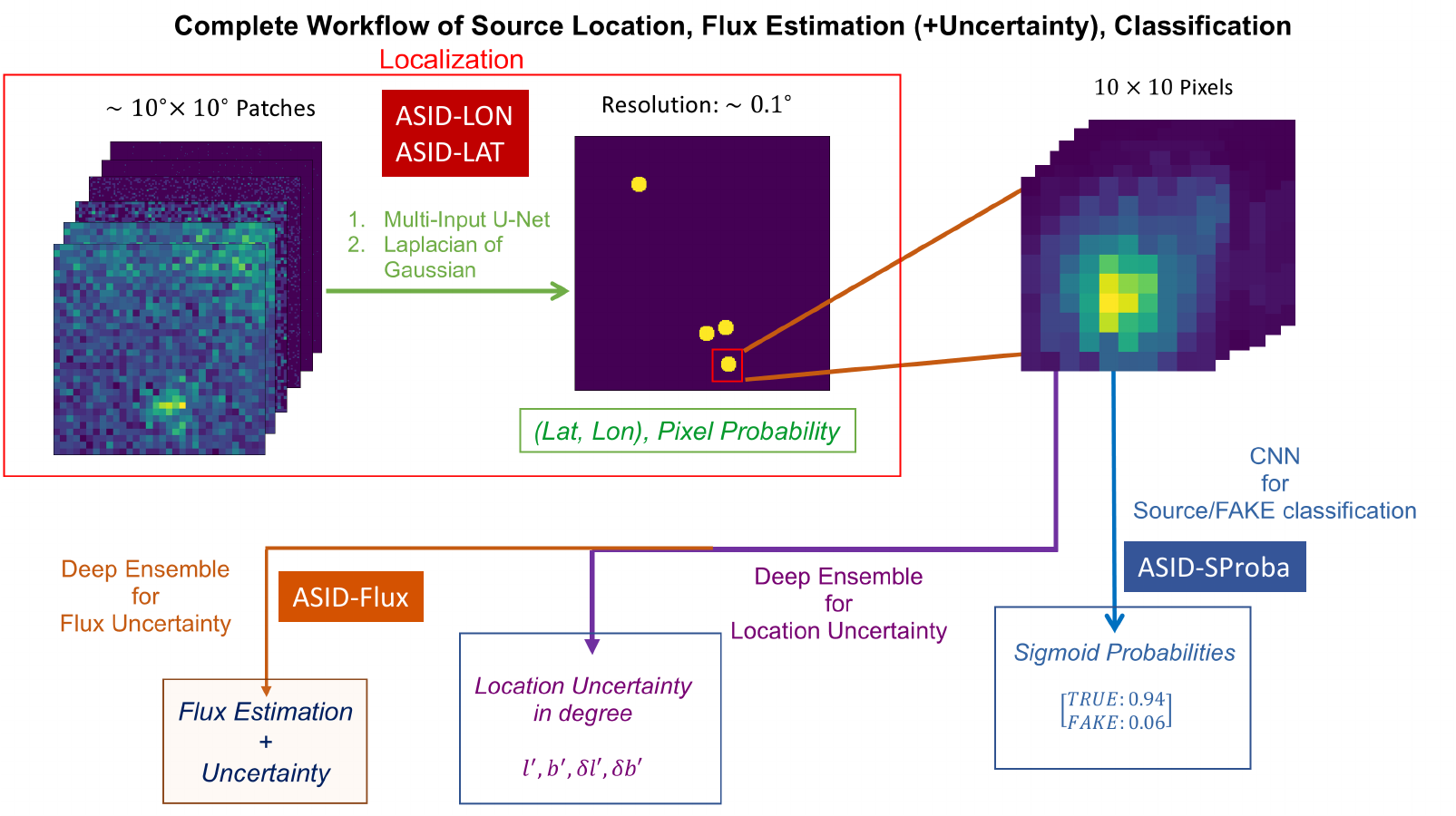}
\caption{Workflow of the updated ASID pipeline consisting in the source detection and localization module, followed by the modules on source classification, position refinement, and flux estimation including uncertainties.}
\label{fig.1}
\end{figure}

We test the updated ASID network on 10 years of simulated \textit{Fermi} data using collaboration software \texttt{Fermitools}\footnote{\url{https://fermi.gsfc.nasa.gov/ssc/data/analysis/software/}}, and population models from the 4FGL-DR2 catalog \cite{Ballet:2020hze}. The simulations are binned into 6 energy bins (from 300 MeV < $E$ < 1 TeV) and in 10 deg $\times$ 10 deg patches. ASID is based on a Multi-Input U-Net architecture for object localization, producing segmented regions around point source centers and followed by a Laplacian of Gaussian method for clustering, which provides the values for longitude and latitude of each detected source. For details on the specifics of the architecture, hyperparameter setups, training details or timings, we refer to our previous work \cite{Panes:2021zig}. This stage is followed by the characterization module, applied to cut-out regions around the predicted sources, and composed of a classification module (VGG-like CNN for binary classification between True and Fake classes), and a flux estimation and location refinement modules (both based on deep ensemble networks). Combining the outputs of the complete model, the final output is an astrophysical catalog.  The updated ASID workflow is depicted in Figure~\ref{fig.1}.

To assess the performance of our pipeline, we plot \textit{Recall} (also known as \textit{Completeness}), and defined as:
\begin{equation}
\label{eq.1}
Recall=\frac{True\; positives}{True\; positives\; +\; False\; negatives},
\end{equation}
versus the true source flux ($F$). In this plot, we observe that the flux sensitivity of ASID is comparable to the 4FGL-DR2 detection threshold and in agreement with the previous application of ASID \cite{Panes:2021zig}, being around $F\approx2\times 10^{-10}\;{\rm cm^{-2}\;s^{-1}}$. To test the robustness against different IEM models, we use the pipeline trained with one model (B1-IEM) and test it against data simulated using a different IEM (B1-IEM), finding that the overall amount of true sources detected is consistent. In both the \textit{Recall} plot and the test with different IEMs, we find the best performance of the pipeline is for the region of latitudes $|b|>20$ deg. Finally, we apply ASID to real \textit{Fermi}-LAT data. We highlight that for 4FGL-DR2 sources detected with significance $\sigma > 20$ and $|b|>20$ deg, we obtain 98\% of associations. 

\section{CTAO: challenges of the new generation of IACTS}
\label{sec:ctao}
The CTAO \cite{CTAConsortium:2010umy} is the next generation of Imaging Air Cherenkov Telescopes (IACTs), operating between 20 GeV and 300 TeV; and serving as the reference facility in its energy domain. \textit{Fermi}-LAT and CTAO yield complementary windows on the gamma-ray sky: \textit{Fermi}-LAT provides whole-sky observations at low energies, while CTAO operates at higher energies through targeted observations. The backgrounds also differ: \textit{Fermi}-LAT observations are dominated by the IEM, whereas CTAO will be primarily limited by instrumental background. With an expected order-of-magnitude improvement in sensitivity over current IACTs\footnote{\url{https://www.ctao.org/for-scientists/performance/}}, CTAO is expected to detect over an order of magnitude more faint gamma-ray sources \cite{CTAConsortium:2023tdz}. Combined with its improved angular resolution, many of these will appear extended, producing overlapping emission regions. Together with differences in background spectrum and morphology, these characteristics underscore the substantial gap between \textit{Fermi}-LAT and CTAO data.

\begin{figure}[h!]
\centering
\includegraphics[width=\linewidth]{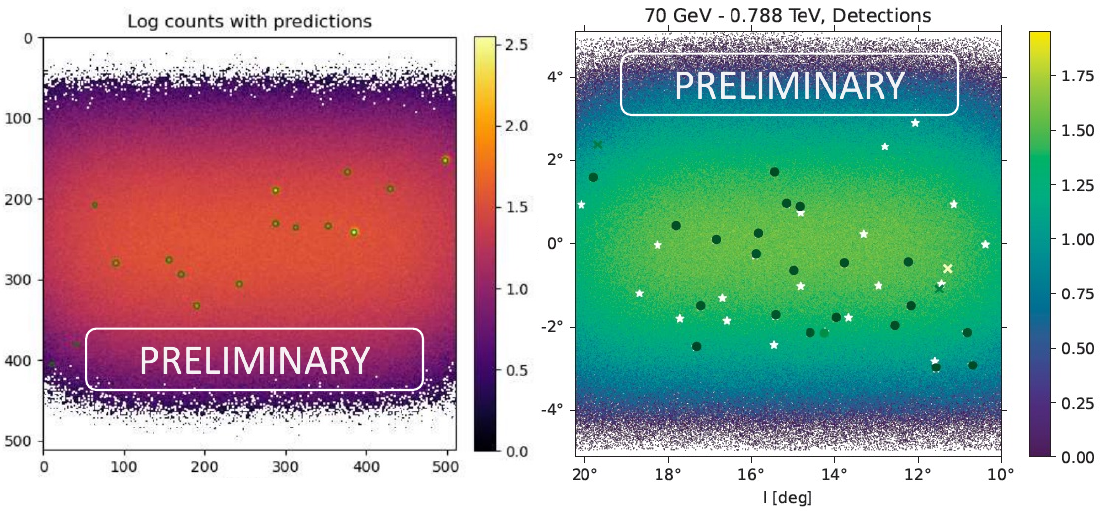}
\caption{Example patches of ASID (\textbf{left panel}) and CeDIRNet (\textbf{right panel}) results using log-scaling on the image. The patches correspond to different realizations. The white stars represent true sources and the green circles the detected sources, and the green crosses the fake positives. The colormaps represent the number of counts.}
\label{fig.2}
\end{figure}

The GP, an extremely complex region of the gamma-ray sky, will be a primary target for the CTAO. To validate ASID performance, we simulate GP data with only point-like sources, as a proof-of-concept, using consortium software \texttt{gammapy} \cite{Gammapy:2023gvb}, in three energy bins (from 70 GeV < $E$ < 100 TeV) and using 10 deg $\times$ 10 deg patches. In preparation for more realistic scenarios, we tested the CeDiRNet algorithm \cite{2024PatRe.15310540T}, a CNN that regresses directions to nearby source centers and is expected to improve performance in crowded regions. Results on example patches are shown in Figure~\ref{fig.2}. On the \textit{Recall} (see Equation~\ref{eq.1}) vs. true source flux plot, both \textit{CeDiRNet} and \textit{ASID with log-scaling patches} show similar results, reaching 90\% recall for sources with flux $F(>{\rm1\;TeV})\approx2\times 10^{-14}\;{\rm cm^{-2}\;s^{-1}}$, also aligned with the results of the standard likelihood approach using \texttt{gammapy}. Both models yield similar performances, providing robustness to the first application of CNN models to CTAO data. Furthermore, they provide similar performance than the standard methods, with the advantage of automatization and improvement of time allocation.

\section{Towards a foundation model: optical data and the latent space}
\label{sec:foundation}

\begin{figure}[h!]
\centering
\includegraphics[width=\linewidth]{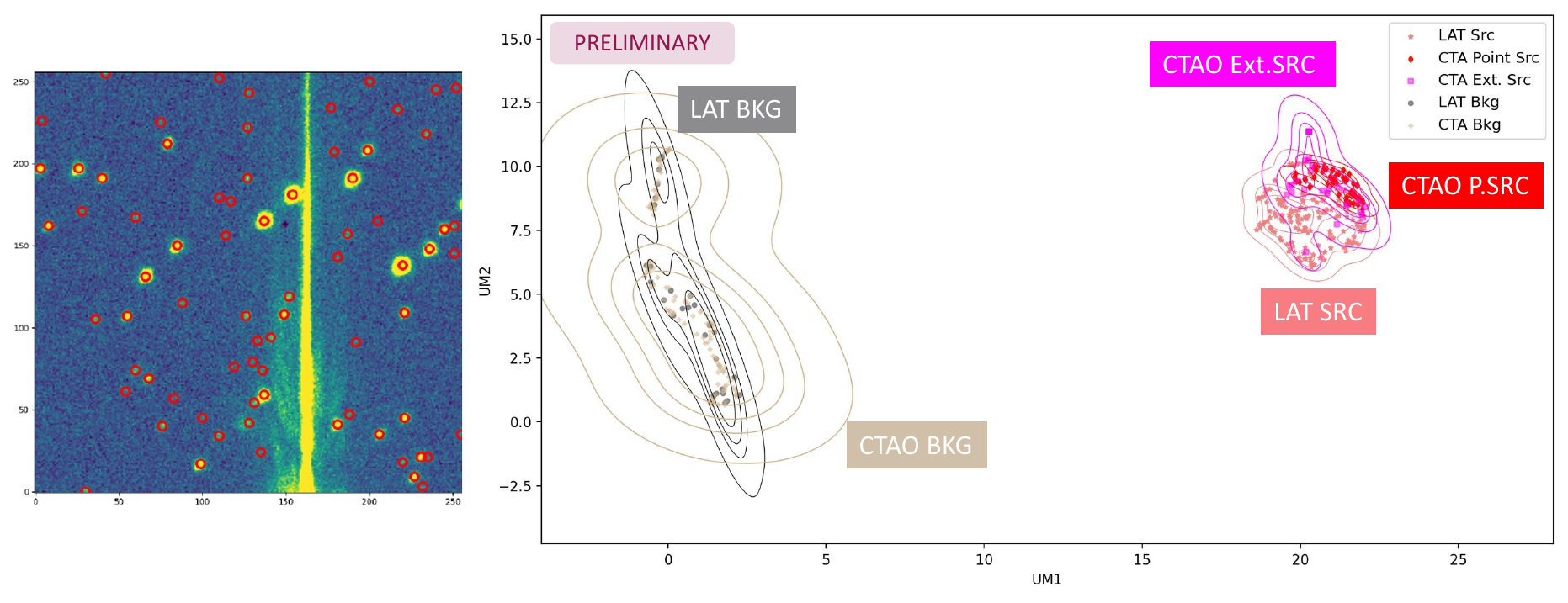}
\caption{Right: Example patch with localized sources by ASID (red circles) superimposed on the optical image in the presence of artifacts (  \cite{Stoppa:2022wta}). Left: Representation of the latent space of the bottleneck of our model. Grey (\textit{Fermi}-LAT) and brown (CTAO) points represent the gamma-ray backgrounds, while red (\textit{Fermi}-LAT) and pink (CTAO) represent the gamma-ray sources.}
\label{fig.3}
\end{figure}

As discussed in Section~\ref{sec:intro}, the integration of multi-wavelength datasets is particularly relevant for catalog construction. Motivated by this, we investigated whether ASID could also be applied beyond the gamma-ray domain. To this end, we trained and tested ASID on MeerLICHT data \cite{2016SPIE.9906E..64B, Stoppa:2022wta}, using images of fields with varying source densities, each divided into 1,681 patches of 256 pix $\times$ 256 pix. An example result is shown in the left panel of Figure~\ref{fig.3}, where ASID not only outperforms standard tools in source detection but also is able to discard artifacts. Furthermore, we examined the impact of image resolution on localization performance by applying ASID to data from the Hubble Space Telescope and the Wide-field Infrared Survey Explorer (WISE).

The adaptability of the ASID framework, as demonstrated in these studies, raises its potential as a building block for a foundation model for astrophysical source detection across wavelengths. As a first step, we explore the latent-space representation of the model’s bottleneck using only gamma-ray data, shown in the right panel of Figure~\ref{fig.3}. The plot reveals two separated clusters: one associated with backgrounds and the other with sources. Notably, sources and background points from both \textit{Fermi}-LAT and CTAO data cluster together, underscoring the feasibility of developing a foundation model, capable of operating on heterogeneous telescope data, for astrophysical source detection and characterization. Further development of this model is ongoing. 

\section{Conclusions}

In these proceedings, we have presented the preliminary results on two applications of a DL-based pipeline for gamma-ray source detection. For \textit{Fermi}-LAT, ASID performance at $|b| < 20^\circ$ and for low-significance sources can be further improved. Future efforts will focus on mitigating the impact of the IEM background and on exploring extraction methods. For the CTAO, the next step will involve implementing a denoising pipeline and moving towards more realistic simulations that include extended sources. The application of ASID to MeerLICHT, Hubble, and WISE data demonstrates its flexibility. As a first step for building a foundation model for astrophysical source detection, we have explored the latent space of the model for gamma-ray data and plan to incorporate multi-wavelength datasets in the near future.

\section*{Acknowledgements}
This publication is co-funded by/ has received funding from/the European Union’s Horizon Europe research and innovation program under the Marie Sklodowska-Curie COFUND Postdoctoral Programme grant agreement No.101081355- SMASH and by/from  the Republic of Slovenia and the European Union from the European Regional Development Fund. Co-funded by the European Union. Views and opinions expressed are however those of the author(s) only and do not necessarily reflect those of the European Union or European Research Executive Agency. Neither the European Union nor the granting authority can be held responsible for them.







\bibliography{SciPost_Example_BiBTeX_File.bib}


\end{document}